\documentclass[aps,prl,twocolumn,10pt,showpacs,superscriptaddress,amsmath,amssymb]{revtex4-1}

\usepackage{graphicx}

\begin{document}

\title{Are Observables necessarily Hermitian?}
\author{Meng-Jun Hu}
\author{Xiao-Min Hu}
\author{Yong-Sheng Zhang}\email{yshzhang@ustc.edu.cn}
\affiliation{Laboratory of Quantum Information, University of Science and Technology of China, Hefei, 230026, China}
\affiliation{Synergetic Innovation Center of Quantum Information and Quantum Physics, University of Science and Technology of China, Hefei, 230026, China}

\date{\today}

\begin{abstract}
Observables are believed that they must be Hermitian in quantum theory. Based on the obviously physical fact that only eigenstates of observable and its corresponding probabilities, i.e., spectrum distribution of observable are actually observed, we argue that observables need not necessarily to be Hermitian. More generally, observables should be reformulated as normal operators including Hermitian operators as a subclass. This reformulation is consistent with the quantum theory currently used and does not change any physical results. The Clauser-Horne-Shimony-Holt (CHSH) inequality is taken as an example to show that our opinion does not conflict with conventional quantum theory  
and gives the same physical results. Reformulation of observables as normal operators not only coincides with the physical facts but also will deepen our understanding of measurement in quantum theory. 
\end{abstract}


\maketitle

Observable is an essential concept in quantum theory. Based on the ``obviously" physical fact that the measured result must be real when we make an observation of some physical quantity, it is widely accepted as a basic postulate in quantum theory that the possible quantum observables must be Hermitian \cite{1}. Since the eigenvalues of Hermitian operators are real, it seems natural to ascribe Hermiticity to observables. If an observable can be a non-Hermitian operator, there seems no experimental way exists to give expectation value of such operator since the spectrum of a non-Hermitian operator is complex in general. With all these convincing facts, the idea that observables must be Hermitian has been deeply rooted in our mind.

However, the realness of spectrum of observables does not imply that they must be Hermitian operators. There exists non-Hermitian operators that satisfy some symmetry conditions have real eigenvalues. It thus not a valid conclusion that observables must be Hermitian based on the so called ``obviously" physical fact. In fact, it is physically admitted the existence of observables that are not Hermitian but possess real spectrum \cite{2,3,4}. More surprisingly, recent research \cite{5} seems to show that quantum theory allows direct measurement of any non-Hermitian operator via weak value introduced by Aharonov {\it et al} \cite{6,7,8}.  
All these facts indicate the widely accepted idea that observables are Hermitian need to be modified  so that it can coincide with recent researches.

The state of quantum object in quantum theory is completely described by a wave function $\psi$ which is totally different from classical theory in which the state of classical object is described by its position and moment $(q,p)$, etc. In order to detect the physical quantity of quantum object, a suitable setup has to be designed to do the measurement. The results shown in the meter reflect the state of the object we have measured. The state of the object will stay unchanged when we repeat the same measurement. These stationary states observed are called the eigenstates of an observable $A$ we measured and can be proved that they are orthogonal to each other \cite{9}. 
The original state of a quantum object is believed in a superposition of these eigenstates $\psi=\sum_{i}\alpha_{i}\psi_{i}$ and the coefficients $\alpha_{i}$ are related with the probabilities $p_{i}=|\alpha_{i}|^{2}$ that we find corresponding eigenstates $\psi_{i}$ via Born rule. 
This is the whole story of postulate of measurement in quantum theory.      
{\it What we actually measured are the eigenstates of an observable $A$ and its corresponding emerging probabilities, i.e., the spectral distribution $\lbrace\psi_{i},p_{i}\rbrace$ of that observable}. The eigenvalues $\lbrace a_{i}\rbrace$ of observable $A$ are not directly observed but rather given to build correspondence with corresponding eigenstates. There seems no fundamental physical requirement that eigenvalues must be real. Even in the classical case, it is the state of  object that is observed. The value of physical quantity make sense only when a suitable standard of metric is settled.  
The requirement of Hermiticity of observables will be conflicted with the above physical facts in some cases. Suppose that there are two Hermitian observables $\hat{C}$ and $\hat{D}$ that are commutative, i.e., $[\hat{C},\hat{D}]=0$. The question is whether the new operator defined as $\hat{F}\equiv\hat{C}+i\hat{D}$ is an observable or not. The answer is negative according to conventional point of view since $\hat{F}$ is not Hermitian, i.e., $\hat{F}^{\dagger}\neq\hat{F}$. However, this is not compatible with the physical fact that the spectral distribution of $\hat{F}$ can be observed experimentally because of commutativity of observables $\hat{C}$ and $\hat{D}$.
For instance, the value $c+id$ can be given to $\hat{F}$ if we obtain a common eigenstate $|\psi\rangle$ of $\hat{C}$ and $\hat{D}$ in a measurement with $\hat{C}|\psi\rangle=c|\psi\rangle$ and $\hat{D}|\psi\rangle=d|\psi\rangle$.
Based on these physical facts that in a measurement only eigenstates are observed experimentally and eigenvalues are given according to the spectrum of operator of the physical quantity , we will argue, in this letter, that observables need not necessarily be Hermitian. More generally, observables should be reformulated as {\it normal operators} that include Hermitian operators as a subclass. We emphasize that this reformulation does not change any physical results in quantum theory.

As an example, we can consider a typical spin measurement of spin one-half particles via Stern-Gerlach experiment \cite{10}. Suppose that the spin one-half particles are well prepared in the same state with a suitable device. After passing through a gradient magnetic field along $z$ direction, two distinct spots are observed in the screen. The two distinct spots indicate that observed particles in possession of two distinct spin states denoted by $|\uparrow\rangle$ and $|\downarrow\rangle$. If the experiment runs enough many times, the probabilities that particle in spin states $|\uparrow\rangle$ and $|\downarrow\rangle$ can be computed by comparing particle numbers in two distinct spin states. When we pick out particles in one of spin states $|\uparrow\rangle$ or $|\downarrow\rangle$, it can be found that particles will still in the same spin state after passing through the same measurement device subsequently. These stationary distinct states $|\uparrow\rangle$ or $|\downarrow\rangle$ are referred as eigenstates of spin $z$ component of particle represented by an operator $\hat{S_{z}}$. The initial state of spin one-half, according to quantum theory, is generally described by
\begin{equation}
|\psi\rangle=\alpha|\uparrow\rangle+\beta|\downarrow\rangle,
\end{equation}
where $| \alpha|^{2}+| \beta|^{2}=1$. The measured probabilities $p_{\uparrow}$ and $p_{\downarrow}$ corresponding to spin eigenstates $|\uparrow\rangle$ and $|\downarrow\rangle$ are related to the initial spin state $|\psi\rangle$ by Born rule that $p_{\uparrow}=| \alpha|^{2},p_{\downarrow}=| \beta|^{2}$. Thus, the only directly observed result is the spectrum distribution $\lbrace |\uparrow\rangle,p_{\uparrow};|\downarrow\rangle,p_{\downarrow}\rbrace$ of observable $\hat{S}_{z}$.

According to our understanding of classical physical reality \cite{11}, the spin $z$ component $\hat{S}_{z}$ must has definite values in its spin eigenstates. These definite values, denoted by $s_{\uparrow}$ and $s_{\downarrow}$ corresponding to spin eigenstates $|\uparrow\rangle$ and $|\downarrow\rangle$, are called eigenvalues of observable $\hat{S}_{z}$. The exact values of $s_{\uparrow}$ and $s_{\downarrow}$ are not directly measured. All we can infer from the observed results is that eigenvalues $s_{\uparrow}$ and $s_{\downarrow}$ must be opposite with the same magnitude. It seems natural for us to endow real numbers to eigenstates $s_{\uparrow}$ and $s_{\downarrow}$. In this case, observable $\hat{S}_{z}$ is naturally Hermitian that $\hat{S}_{z}^{\dagger}=\hat{S}_{z}$. However, there seems no physical constraint that complex value can not exist since the realness of eigenvalues is somewhat mathematically artificial. If, instead of real numbers, complex numbers are given to eigenvalues, observable must be viewed as an Hermitian operator will make no sense.

People may argue that complex physical quantity is not physical and can never be observed and it seems that one can not imagine the expectation value of observable $\hat{S}_{z}$ to be complex. According to this point of view, the real physical quantity is not physical too since only spin eigenstates $\lbrace|\uparrow\rangle,|\downarrow\rangle\rbrace$ and their probability distribution $\lbrace p_{\uparrow},p_{\downarrow}\rbrace$ are observed, e.g., in the Stern-Gerlach experiment. The value of a physical quantity makes sense only when a standard of metric is settled and physics behind that value is clear to us. The value itself is not physical since it may be changed in different standard of metric. The similar discussions are also applicable to possible complex expectation value of observable. In addition, for any operator has the form of $\hat{O}=a|\uparrow\rangle\langle\uparrow|+b|\downarrow\rangle\langle\downarrow|$ with eigenstates $|\uparrow\rangle$ and $|\downarrow\rangle$, the measurement can be the same as the Stern-Gerlach method. The only difference is just the numerical difference of coefficients $a$ and $b$ but not the physical device.

 The expectation value of any observable $\hat{A}$ in a state $|\psi\rangle$ can be understood as \cite{12}
\begin{equation}
\langle\hat{A}\rangle=\langle\psi|\varphi\rangle,  
|\varphi\rangle\equiv\hat{A}|\psi\rangle.
\end{equation}
Its evolution is completely determined by commutator $[\hat{A},\hat{H}]$ and described by the Heisenberg equation
\begin{equation}
\dfrac{d}{dt}\langle\hat{A}\rangle=\dfrac{1}{i\hbar}[\hat{A},\hat{H}].
\end{equation}
Since quantum commutation relations are independent on whether observables are Hermitian or not, all the physical results remain unchanged. In fact, observables can be viewed as positive operator valued (POV) measures has been proposed by Busch {\it et al} based on physical facts that only the spectrum distribution of an observable is actually observed \cite{13,14}.

The proper physical requirement for an observable should be that its eigenstates must be orthogonal. Based on above discussions, the opinion that observables are viewed as Hermitian operators need to be extended to more general case. As only orthogonal eigenstates actually observed are physical significant, observables should be reformulated as {\it normal operators}. The normality of observable reflects the more fundamental physical requirement that eigenstates of an observable must be mutually orthogonal. It is known that an operator $\hat{B}$ is said to be normal if $\hat{B}$ commutes with its adjoint, that is \cite{15} 
\begin{equation}
\hat{B}^{\dagger}\hat{B}=\hat{B}\hat{B}^{\dagger}.
\end{equation}
Hermitian operators obviously satisfy above definition and compose a subclass of normal operators. The class of normal operators is closed under unitary transformation and any normal operator $\hat{B}$ can be unitarily diagonalizable. If a normal operator $\hat{B}$ represents an observable which has $n$ eigenstates $\lbrace|\lambda_{i}\rangle\rbrace$, then we have 
\begin{equation}
B=U\Lambda U^{\dagger},
\end{equation}
where $\Lambda\equiv\mathrm{diag}(\lambda_{1},\lambda_{2},...,\lambda_{n})$ is a diagonal matrix with $\lbrace\lambda_{i}\rbrace$ are eigenvalues of observable $\hat{B}$. For any eigenstate $|\lambda_{i}\rangle$, it is obvious that
\begin{equation}
\hat{B}|\lambda_{i}\rangle=\lambda_{i}|\lambda_{i}\rangle.
\end{equation}
Any observable should be represented by a normal operator. Conversely, any normal operator can also represent an observable since that eigenstates of a normal operator are orthogonal. Normality is thus a necessary and sufficient condition of observable.

Reformulation of observables as normal operators not only coincides with physical facts that only spectrum distribution of an observable is observed but also can clarify some confusions caused by the requirement of Hermiticity. For above example that $\hat{F}\equiv\hat{C}+i\hat{D}$ with $\hat{C}$ and $\hat{D}$ are Hermitian and commutative operators. The question arises that whether $\hat{F}$ represents an observable. 
Since $\hat{F}^{\dagger}=\hat{C}-i\hat{D}\neq\hat{F}$, the answer is negative based on conventional point of view. However, it is conflicted with physical fact that the eigenstates of $\hat{F}$ and its probability distribution can be observed experimentally because of commutativity of observables $\hat{C}$ and $\hat{D}$. The problem will be automatically solved if we reformulate observable as normal operator. In this new perspective, operator $\hat{F}$ also satisfies normality which can be seen by a simple calculation
\begin{equation}
\hat{F}^{\dagger}\hat{F}=\hat{C}^{\dagger}\hat{C}+i \lbrace\hat{C}^{\dagger}\hat{D}-\hat{D}^{\dagger}\hat{C} \rbrace+\hat{D}^{\dagger}\hat{D}=\hat{F}\hat{F}^{\dagger}.
\end{equation}
The normality of operator $\hat{F}$ is guaranteed by commutation condition $[\hat{C},\hat{D}]=0$. The operator $\hat{F}$ thus indeed represents an observable.

Another example is phase-difference observable \cite{16,17,18}. If the restriction of Hermiticity of observable is relaxed, the operator $\mathrm{exp}(i\hat{P}_{12})$ can be directly used, instead of Hermitian phase-difference operator $\hat{P}_{12}\equiv\hat{\phi_{1}}-\hat{\phi_{2}}$.

To see more clearly that this reformulation does not violate any physical results, we take Bell inequality \cite{19} as an example.  For simplicity, we focus on most commonly discussed Bell inequality, that is Clauser-Horne-Shimony-Holt (CHSH) inequality \cite{20}. The CHSH inequality states that the absolute value of a combination of four correlations in any local-hidden theory is bounded by $2$. Moreover, the combination of quantum correlations appearing in CHSH inequality is bounded by $2\sqrt{2}$ that called Cirel'son bound \cite{21}. Neither classical bound of $2$ nor Cirel'son bound of $2\sqrt{2}$ will be shown  to be changed within our new perspective that observable need not necessarily be Hermitian.
\begin{figure}[tbp]
\centering
\includegraphics[scale=0.5]{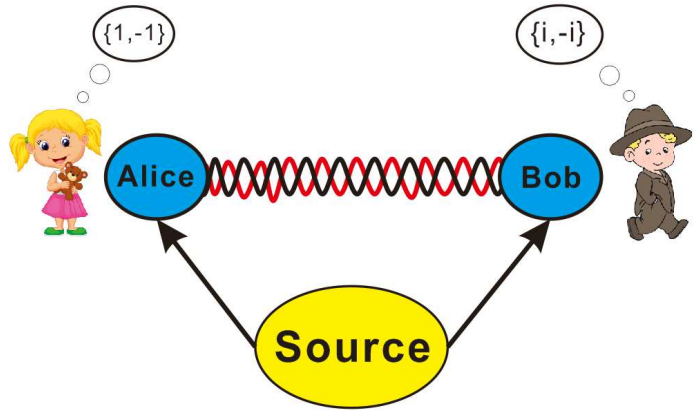}
\caption{(color online)  Two particles $j$ and $k$ are sent to Alice and Bob respectively who are spacelike separated. Only two physical observables $A_{1}$ and $A_{2}$ ($B_{1}$ and $B_{2}$) are available  for Alice (Bob) to perform local measurement on particle $j$ ($k$). For each observable there are two opposite results shown in observer's pointer. Alice, who is a conventional person, gives $1$ or $-1$ to her measurement results $a_{1},a_{2}\in\lbrace 1,-1\rbrace$, while Bob, who is a creative person, gives $i$ or$-i$ to his measurement results $b_{1},b_{2}\in\lbrace i,-i\rbrace$. Neither classical bound of $2$ nor Cirel'son bound of $2\sqrt{2}$ of CHSH inequality will be shown changed. }
\end{figure}

Suppose that two particles $j$ and $k$ of a composite system are sent to Alice and Bob respectively who are spacelike separated. Let $A_{1}$ and $A_{2}$ ($B_{1}$ and $B_{2}$) be physical observables referring to local measurement on particle $j$ ($k$) by Alice (Bob). There are only two opposite results with the same magnitude shown in observer's pointer for each observable. Suppose that Alice and Bob have two different standards of metric. Alice, who is a conventional person, gives $1$ or $-1$ to her measurement results $a_{1},a_{2}\in\lbrace 1,-1\rbrace$, while Bob, who is a creative person, gives $i$ or $-i$ to his measurement results $b_{1},b_{2}\in\lbrace i,-i\rbrace$.
The correlation $C(A_{1},B_{1})$ of observables $A_{1}$ and $B_{1}$ is defined as
\begin{equation}
\begin{split}
C(A_{1},B_{1})=iP_{A_{1}B_{1}}(1,i)-iP_{A_{1}B_{1}}(1,-i) \\
-iP_{A_{1}B_{2}}(-1,i)+iP_{A_{1}B_{1}}(-1,-i),
\end{split}
\end{equation}
where $P_{A_{1}B_{1}}(1,-i)$ represents the joint probability of obtaining $a_{1}=1$ and $b_{1}=-i$. Correlations $C(A_{1},B_{2})$, $C(A_{2},B_{1})$ and $C(A_{2},B_{2})$ can be defined similarly. In the conventional situation, observables $A_{1}$ and $A_{2}$ ($B_{1}$ and $B_{2}$) are believed to be Hermitian and can only take values $1$ or $-1$. In this case, it can be proved, according to any local-hidden theory, that the absolute value of a particular combination of correlations is bounded by $2$. This result remains unchanged in our situation, that is 
\begin{equation}
|C(A_{1},B_{1})+C(A_{1},B_{2})+C(A_{2},B_{1})-C(A_{2},B_{2})|\leq2.
\end{equation}

The bound of $2$ can be simply derived \cite{22}. For a separated system, the observables $A_{1}$ and $A_{2}$ ($B_{1}$ and $B_{2}$) have predefined values $a_{1},a_{2}\in\lbrace 1,-1\rbrace$ ($b_{1},b_{2}\in\lbrace i,-i\rbrace$) in a local-hidden theory. Therefore, for a separated system the combination of correlations in the left hand side of Eq. (9) becomes
\begin{equation}
\begin{split}
S=a_{1}b_{1}+a_{1}b_{2}+a_{2}b_{1}-a_{2}b_{2} \\
=b_{1}(a_{1}+a_{2})+b_{2}(a_{1}-a_{2}).
\end{split}
\end{equation}

It is obvious that $S$ is either $2i$ or $-2i$ since one of expressions $a_{1}+a_{2}$ and $a_{1}-a_{2}$ must be zero and the other is either $2$ or $-2$. Thus, the absolute value of the combination of correlations is still bounded by $2$.

In the quantum case, the state of a two-particle system is described by a state vector $|\psi\rangle$ and the quantum correlation of $A_{1}$ and $B_{1}$ is defined as
\begin{equation}
C_{Q}(A_{1},B_{1})=\langle\psi|\hat{A_{1}}\hat{B_{1}}|\psi\rangle,
\end{equation}
where operators $\hat{A_{1}}$ and $\hat{B_{1}}$ represent observables $A_{1}$ and $B_{1}$ respectively. Note that we do not require $\hat{A_{1}}$ and $\hat{B_{1}}$ to be self-adjoint operators here, which is different from the conventional case. Quantum correlations $C_{Q}(A_{1},B_{2})$, $C_{Q}(A_{2},B_{1})$ and $C_{Q}(A_{2},B_{2})$ can be defined similarly. If observables $A_{1}$, $A_{2}$, $B_{1}$, $B_{2}$ and the state $|\psi\rangle$ of the two-particle system  are properly settled, the same combination of quantum correlations can violate the CHSH inequality which means no local-hidden theory can reproduce the predictions of quantum theory. 

The combination of quantum correlations obeys some limitations too. Cirel'son proved that the absolute value of the combination of quantum correlations that is similar to those appearing in the CHSH inequality $(9)$ in a two-particle system is bounded by $2\sqrt{2}$ \cite{21}
\begin{equation}
 |C_{Q}(A_{1},B_{1})+C_{Q}(A_{1},B_{2})+C_{Q}(A_{2},B_{1})-C_{Q}(A_{2},B_{2})|\leq 2\sqrt{2}.
 \end{equation} 
We will prove, in the following, that Cirel'son bound does not change whether or not observables are Hermitian. 

To derive the Cirel'son bound, the new operator which has the same structure of combination of correlations is constructed as
\begin{equation}
\hat{Z}=\hat{A_{1}}\hat{B_{1}}+\hat{A_{1}}\hat{B_{2}}+\hat{A_{2}}\hat{B_{1}}-\hat{A_{2}}\hat{B_{2}}.
\end{equation}
Calculating the operator $\hat{Z}^{\dagger}\hat{Z}$ gives
\begin{equation}
\begin{split}
\hat{Z}^{\dagger}\hat{Z}=\hat{A_{1}}^{\dagger}\hat{A_{1}}\hat{B_{1}}^{\dagger}\hat{B_{1}}+\hat{A_{1}}^{\dagger}\hat{A_{1}}\hat{B_{2}}^{\dagger}\hat{B_{2}} \\
+\hat{A_{2}}^{\dagger}\hat{A_{2}}\hat{B_{1}}^{\dagger}\hat{B_{1}}+\hat{A_{2}}^{\dagger}\hat{A_{2}}\hat{B_{2}}^{\dagger}\hat{B_{2}} \\
+(\hat{A_{1}}^{\dagger}\hat{A_{1}}-\hat{A_{2}}^{\dagger}\hat{A_{2}})(\hat{B_{1}^{\dagger}}\hat{B_{2}}+\hat{B_{2}^{\dagger}}\hat{B_{1}}) \\
+(\hat{B_{1}}^{\dagger}\hat{B_{1}}-\hat{B_{2}}^{\dagger}\hat{B_{2}})(\hat{A_{1}^{\dagger}}\hat{A_{2}}+\hat{A_{2}^{\dagger}}\hat{A_{1}}) \\
+(\hat{A_{1}^{\dagger}}\hat{A_{2}}-\hat{A_{2}^{\dagger}}\hat{A_{1}})(\hat{B_{2}^{\dagger}}\hat{B_{1}}-\hat{B_{1}^{\dagger}}\hat{B_{2}}).
\end{split}
\end{equation}
For Hermitian observables of $A_{i}$ and $B_{i}$, $\hat{A_{1}}^{\dagger}\hat{A_{1}}=\hat{A_{2}}^{\dagger}\hat{A_{2}}=\hat{B_{1}}^{\dagger}\hat{B_{1}}=\hat{B_{2}}^{\dagger}\hat{B_{2}}=I$ with $I$ is identity operator and Eq. (14) is simplified into
\begin{equation}
\hat{Z}^{2}=4I-[\hat{A_{1}},\hat{A_{2}}][\hat{B_{1}},\hat{B_{2}}].
\end{equation}
Since for all bounded operator $\hat{M}$ and $\hat{J}$,
\begin{equation}
||[\hat{M},\hat{J}]||\leq||\hat{M}\hat{J}||+||\hat{J}\hat{M}||\leq 2||\hat{M}||\cdot||\hat{J}||,
\end{equation}
it can be shown
\begin{equation}
||\hat{Z}||^{2}\leq||\hat{Z}^{2}||\leq 4+4||\hat{A_{1}}||\cdot||\hat{A_{2}}||\cdot||\hat{B_{1}}||\cdot||\hat{B_{2}}||=8,
\end{equation}
so that $||\hat{Z}||\leq 2\sqrt{2}$.

In fact, the bound $2\sqrt{2}$ is independent of Hermiticity of observables. If $\hat{A_{i}}$ and $\hat{B_{i}} (i=1,2)$ are restricted by $||\hat{A_{i}}||=||\hat{B_{i}}||=I$, Alice and Bob can select any pair of number $\lbrace e^{i\phi_{1}},e^{i\phi_{2}}\rbrace$ corresponding to their measurement results and surprisingly the physical results still stay unchanged since in this case we have
\begin{equation}
||\hat{Z}^{\dagger}\hat{Z}||=||4I+(\hat{A_{1}^{\dagger}}\hat{A_{2}}-\hat{A_{2}^{\dagger}}\hat{A_{1}})(\hat{B_{2}^{\dagger}}\hat{B_{1}}-\hat{B_{1}^{\dagger}}\hat{B_{2}})||\leq 8.
\end{equation}
Therefore, Cirel'son bound of $2\sqrt{2}$ is hold.

In conclusion, based on physical fact that only eigenstates of observable and its corresponding probabilities distribution are actually observed in experiment, we conclude that observables need not necessarily be Hermitian but rather should be reformulated as normal operators including Hermitian operators as a subclass. Bell inequality is taken as a representative example to show that our work is not conflicted with physical results of conventional quantum theory. Neither classical bound of $2$ nor Cirel'son bound of $2\sqrt{2}$ in CHSH inequality is changed, which implies that physical results are independent of whether or not observables are Hermitian. Reformulation of observables as normal operators not only coincides with the physical facts but also will deepen our understanding of measurement in quantum theory.

This work is supported by National Natural Science Foundation of China (No. 61275122, No. 61590932), National Fundamental Research Program of China and Strategic Priority Research Program (B) of CAS (No. XDB01030200).


\begin{thebibliography}{99}

\bibitem{1} P. A. M. Dirac, {\it The Principles of Quantum Mechanics} (Clarendon, Oxford, 1958).

\bibitem{2} C. M. Bender and S. Boettcher, Phys. Rev. Lett. {\bf 80}, 5243 (1998).

\bibitem{3} C. M. Bender, D. C. Brody, and H. F. Jones, Phys. Rev. Lett. {\bf 89}, 270401 (2002); C. M. Bender, D. C. Brody, and H. F. Jones, Phys. Rev. Lett. {\bf 92}, 119902 (2004).

\bibitem{4} C. M. Bender, Rep. Prog. Phys. {\bf 70}, 947-1018 (2007).

\bibitem{5} A. K. Pati, U. Singh, and U. Sinha, Phys. Rev. A {\bf 92}, 052120 (2015).

\bibitem{6} Y. Aharonov, D. Z. Albert, and L. Vaidman, Phys. Rev. Lett. {\bf 60}, 1351 (1988).

\bibitem{7} Y. Aharonov and L. Vaidman, Phys. Rev. A {\bf 41}, 11 (1990).

\bibitem{8} Y. Aharonov and A. Botero, Phys. Rev. A {\bf 72}, 052111 (2005).

\bibitem{9} W. H. Zurek, arXiv:0707.2832v1 [quant-ph] (2007).

\bibitem{10} J. J. Sakurai, San Fu Tuan. {\it Modern Quantum Mechanics, Revised Edition} (Addison-Wesley, Boston, 1994).

\bibitem{11} A. Einstein, B. Podolsky, and N. Rosen, Phys. Rev. {\bf 47}, 777 (1935).

\bibitem{12} R. P. Feynman, R. B. Leighton, and M. Stands, {\it The Feynman Lectures on Physics}, Vol.3 (Addison-Wesley, Boston, 1964).

\bibitem{13} P. Busch, M. Grabowski, and P. J. Lahti, {\it Operational Quantum Physics} (Springer-Verlag, Berlin, 1995).

\bibitem{14} P. Busch, P. Lahti, and R. F. Werner, Rev. Mod. Phys. {\bf 86}, 1261 (2014).

\bibitem{15} R. A. Horn and C. R. Johnson, {\it Matrix Analysis, Second Edition} (Cambridge University Press, Cambridge, 2013).

\bibitem{16} R. Lynch, Phys. Rep. {\bf 256}, 367-436 (1995).

\bibitem{17} J. R. Torgerson and L. Mandel, Phys. Rev. Lett. {\bf 76}, 3939 (1996).

\bibitem{18} S. X. Yu, Phys. Rev. Lett. {\bf 79}, 780 (1997).

\bibitem{19} J. S. Bell, Physics(Long Island City, N.Y.) {\bf 1}, 195 (1964).

\bibitem{20} J. F. Clauser, M. A. Horne, A. Shimony, and R. A. Holt, Phys. Rev. Lett. {\bf 23}, 880 (1969). 

\bibitem{21} B. S. Cirel'son, Lett. Math. Phys. {\bf 4}, 93 (1980).

\bibitem{22} A. Cabello, Phys. Rev. Lett. {\bf 88}, 060403 (2002).

\end{thebibliography}
\end{document}